\documentclass[a4paper,10pt]{article}

\usepackage[utf8]{inputenc}
\usepackage[spanish,english]{babel}
\usepackage{graphicx}
\usepackage{amssymb}
\usepackage{amsmath}
\usepackage{hyperref}
\usepackage{color}
\usepackage{latexsym,dsfont,mathtools}
\usepackage{theorem,titletoc,titlesec}
\usepackage{enumitem}
\usepackage{tikz}
\usepackage{tabularx}
\usetikzlibrary{matrix}
\usepackage{adjustbox}
\newtheorem{theorem}{Theorem}[section]
\newtheorem{corollary}{Corollary}[section]
\newtheorem{proposition}{Proposition}[section]
\newtheorem{lemma}{Lemma}[section]
\newtheorem{example}{Example}[section]
\newtheorem{remark}{Remark}[section]

\usepackage{multirow}
\usepackage{tabularx}
\usepackage{xcolor}

\newcommand{\Z}{\mathbb{Z}}

\newcommand{\zero}{{\mathbf{0}}}
\newcommand{\one}{{\mathbf{1}}}
\newcommand{\p}{{\mathbf{p}}}

\newcommand{\C}{{\cal C}}

\newcommand{\wt}{{\rm wt}}

\newcommand{\uu}{\mathbf{u}}
\newcommand{\vv}{\mathbf{v}}

\newcommand{\cS}{{\cal S}}

\newcommand{\cH}{{\cal H}}

\begin{document}

\title{Construction and Linearity of  $\Z_p\Z_{p^2}$-Linear \\ Generalized Hadamard Codes 
\thanks{This work has been partially supported by the Spanish MINECO under Grant PID2019-104664GB-I00
(AEI / 10.13039/501100011033) and by Catalan AGAUR scholarship 2020 FI SDUR 00475.}
\thanks{The material in this paper was presented in part at the 12th International Workshop on Coding and Cryptography (WCC 2022), 7-11 March 2022 - Virtual.}
}

\author{Dipak K. Bhunia, Cristina Fern\'andez-C\'ordoba, Merc\`e Villanueva \thanks{The authors are with the Department of Information and Communications
Engineering, Universitat Aut\`{o}noma de Barcelona, 08193 Cerdanyola del Vall\`{e}s, Spain.}}



\maketitle

\begin{abstract}
The $\Z_p\Z_{p^2}$-additive codes are subgroups of $\Z_p^{\alpha_1} \times \Z_{p^2}^{\alpha_2}$, and can be seen as linear codes over $\Z_p$ when $\alpha_2=0$,  $\Z_{p^2}$-additive codes when $\alpha_1=0$, or $\Z_2\Z_4$-additive codes when $p=2$. A $\Z_p\Z_{p^2}$-linear generalized Hadamard (GH) code is a GH code over $\Z_p$ which is the Gray map image of a $\Z_p\Z_{p^2}$-additive code.  In this paper, we generalize some known results for $\Z_p\Z_{p^2}$-linear GH codes with $p=2$ to any $p\geq 3$ prime when $\alpha_1 \neq 0$. First, we give a recursive construction of $\Z_p\Z_{p^2}$-additive GH codes of type $(\alpha_1,\alpha_2;t_1,t_2)$ with $t_1,t_2\geq 1$. Then, we show  for which types the corresponding $\Z_p\Z_{p^2}$-linear GH codes  are  non-linear over $\Z_p$. Finally, according to some computational results, we see that, unlike $\Z_4$-linear GH codes, when $p\geq 3$ prime, the $\Z_{p^2}$-linear GH codes are not included in the family of $\Z_p\Z_{p^2}$-linear GH codes with $\alpha_1\not =0$.
\end{abstract}




\section{Introduction}
Let $\Z_p$ and $\Z_{p^2}$ be the ring of integers modulo $p$ and $p^2$, respectively, where $p$ is a prime. Let $\Z_p^n$ and $\Z_{p^2}^n$ denote the set of all $n$-tuples over $\Z_p$ and $\Z_{p^2}$, respectively. In this paper,
the elements of $\Z_p^n$ and $\Z^n_{p^2}$ will be called vectors of length $n$. 
The order of a vector $\mathbf u$ over $\Z_{p^2}$, denoted by $o(\mathbf{u})$, is the smallest positive integer $m$ such that $m \mathbf{u} =(0,\ldots,0)$.

A code over $\Z_p$ of length $n$ is a nonempty subset of $\Z_p^n$,
and it is linear if it is a subspace of $\Z_{p}^n$. Similarly, a nonempty
subset of $\Z_{p^2}^n$ is a $\Z_{p^2}$-additive if it is a subgroup of $\Z_{p^2}^n$. A $\Z_p\Z_{p^2}$-additive code is a subgroup of $\Z_p^{\alpha_1} \times \Z_{p^2}^{\alpha_2}$. 
Note that a $\Z_p\Z_{p^2}$-additive code is a linear code over $\Z_p$ when $\alpha_2=0$,  a $\Z_{p^2}$-additive code when $\alpha_1=0$, or a $\Z_2\Z_4$-additive code when $p=2$. 


The Hamming weight of a vector $\textbf{u}\in\Z_{p}^n$, denoted by $\wt_H(\textbf{u})$, is
the number of nonzero coordinates of $\textbf{u}$. The Hamming distance of two
vectors $\textbf{u},\textbf{v}\in\Z_{p}^n$, denoted by $d_H(\textbf{u},\textbf{v})$, is the number of
coordinates in which they differ.  Note that $d_H(\textbf{u},\textbf{v})=\wt_H(\textbf{u}-\textbf{v})$. The minimum distance of a code $C$ over $\Z_p$ is $d(C)=\min \{ d_H(\textbf{u},\textbf{v}) : \textbf{u},\textbf{v} \in C, \textbf{u} \not = \textbf{v}  \}$. 

In \cite{Sole}, a Gray map  from $\Z_4$ to $\Z_2^2$ is defined as
$\phi(0)=(0,0)$, $\phi(1)=(0,1)$, $\phi(2)=(1,1)$ and $\phi(3)=(1,0)$. There exist different generalizations of this Gray map, which go from $\Z_{2^s}$ to
$\Z_2^{2^{s-1}}$ \cite{Carlet,Codes2k,dougherty,Nechaev,Krotov:2007}.

The one given in \cite{Nechaev} can be defined in terms of the elements of a Hadamard code \cite{Krotov:2007}, and Carlet's Gray map \cite{Carlet} is a particular case of the one given in \cite{Krotov:2007} 
satisfying $\sum \lambda_i \phi(2^i) =\phi(\sum \lambda_i 2^i)$ \cite{KernelZ2s}. 
In this paper, we focus on a generalization of Carlet's Gray map, from $\Z_{p^s}$ to $\Z_p^{p^{s-1}}$, which is also a particular case of the one given in \cite{ShiKrotov2019}. Specifically, 
\begin{equation}
\begin{split}\label{eq:GrayMapCarlet}
\phi: & \ \Z_{p^2} \longrightarrow \Z_p^p \\
      & \ u  \mapsto (u_0,u_1)M,
\end{split}
\end{equation}
where $u\in\Z_{p^2}$; $[u_0,u_1]_p$ is the $p$-ary expansion of $u$, that is $u=u_0 + u_1p$ with $u_0, u_1 \in \Z_p$; and $M$ is the following matrix of size $2 \times p$:
$$\left(\begin{array}{ccccc}
0 & 1 & 2 &\cdots  & p-1 \\
1 & 1 & 1 &\cdots  & 1 \\
\end{array}\right).$$
Let $\Phi:\Z_p^{\alpha_1} \times \Z_{p^2}^{\alpha_2} \rightarrow\Z_p^n$, where $n=\alpha_1+p\alpha_2$, be an extension of the Gray map $\phi$ given by 
$$
\Phi(\mathbf{x}\mid \mathbf{y})=(\mathbf{x}, \phi(y_1), \dots, \phi(y_{\alpha_2})),
$$
for any $\mathbf{x} \in \Z_p^{\alpha_1}$ and  $\mathbf{y}=(y_1,\dots,y_{\alpha_2}) \in \Z_{p^2}^{\alpha_2}$.

Let $\C$ be a $\Z_p\Z_{p^2}$-additive code over $\Z_p^{\alpha_1} \times \Z_{p^2}^{\alpha_2}$. We say that its Gray map image
$C=\Phi(\C)$ is a $\Z_p\Z_{p^2}$-linear code of length $\alpha_1 +p\alpha_2$.
Since $\C$ is isomorphic to a subgroup of
$\Z_{p^2}^{\alpha_1+\alpha_2}$, it is also isomorphic to an abelian structure
$\Z_{p^2}^{t_1}\times\Z_p^{t_2}$, and we say that $\C$, or equivalently
$C=\Phi(\C)$, is of type $(\alpha_1, \alpha_2 ;t_1,t_{2})$.
Note that $|\C|=p^{2t_1+t_2}$. Unlike linear codes over finite fields, linear codes over rings do not have a basis, but there exists a generator matrix for these codes having minimum number of rows, that is, $t_1+t_2$ rows.

A generalized Hadamard $(GH)$ matrix $H(p,\lambda) = (h_{i j})$ of order $n = p\lambda$ over $\Z_p$ is a $p\lambda \times p\lambda$ matrix with entries from $\Z_p$ with the property that for every $i, j$, $1 \leq i < j \leq p\lambda,$ each of the multisets $\{h_{is}- h_{js} : 1 \leq s \leq p\lambda\}$ contains every element of $\Z_p$ exactly $\lambda$ times \cite{jungnickel1979}. 
An ordinary Hadamard matrix of order $4\mu$ corresponds to  a $GH$ matrix $H(2,\lambda)$ over $\Z_2$, where $\lambda = 2\mu$ \cite{Key}. 
Two $GH$ matrices $H_1$ and $H_2$ of order $n$ are said to be equivalent if one can be obtained from the other by a permutation of the rows and columns and adding the same element of $\Z_p$ to all the coordinates in a row or in a column. 

We can always change the first row and column of a $GH$ matrix into zeros and we obtain an equivalent $GH$ matrix which is called normalized. From a normalized GH matrix $H$, we denote by $F_H$ the code consisting of the rows of $H$, and $C_H= \bigcup_{\alpha \in\Z_p} (F_H + \alpha \textbf{1})$,
where $F_H + \alpha \textbf{1} = \{\textbf{h} + \alpha \textbf{1} : \textbf{h} \in F_H\}$ and $\textbf{1}$ denotes the
all-one vector. The code $C_H$ over $\Z_p$ is called generalized
Hadamard $(GH)$ code \cite{dougherty2015ranks}. Note that $C_H$ is generally a non-linear code over $\Z_p$. Moreover, if it is of length $N$, it has $pN$ codewords and minimum distance $N(p-1)/p$.

The $\Z_p\Z_{p^2}$-additive codes such that after the Gray map $\Phi$ give
GH codes are called $\Z_p\Z_{p^2}$-additive GH codes and the
corresponding images are called $\Z_p\Z_{p^2}$-linear GH codes. The classification of $\Z_2\Z_4$-linear GH codes of length $2^t$ with $\alpha_1=0$ and $\alpha_1\not =0$ is given in \cite{Kro:2001:Z4_Had_Perf,PRV06}, showing that there are $\lfloor
(t-1)/2\rfloor$  and $\lfloor
t/2\rfloor$ such non-equivalent codes, respectively.
 
Moreover, in \cite{KV2015}, it is shown that each $\Z_2\Z_4$-linear GH code with $\alpha_1 =0$ is  equivalent to a  $\Z_2\Z_4$-linear GH code with $\alpha_1 \not =0$, so indeed there are only $\lfloor
t/2\rfloor$ non-equivalent $\Z_2\Z_4$-linear GH codes of length $2^t$.
Later, in \cite{KernelZ2s,HadamardZps,fernandez2019mathbb,EquivZ2s}, an iterative construction for $\Z_{p^s}$-linear GH codes is described, the linearity is established, and a partial classification 
is obtained,
giving the exact amount of non-equivalent non-linear such codes for some parameters. 

This paper is focused on $\Z_p\Z_{p^2}$-linear GH codes with $\alpha_1\not =0$ and $p\geq 3$ prime, generalizing some results given for $p=2$ in \cite{PRV06,RSV08} related to the construction and linearity of such codes. For $p=3$ and $2\leq t\leq 8$, these codes are compared with the $\Z_p\Z_{p^2}$-linear GH codes of length $p^t$ with $\alpha_1=0$ studied in \cite{HadamardZps}.
This paper is organized as follows.
In Section \ref{Sec:GrayMap}, we recall the definition of the Gray map considered in this paper and some of its properties.
In Section \ref{Sec:construction}, we describe the construction of  $\Z_p\Z_{p^2}$-linear GH codes of type $(\alpha_1,\alpha_2;t_1,t_2)$ with $\alpha_1\not =0$.
In Section \ref{Sec:Linearity}, we establish for which types these codes are linear.
Finally, in Section \ref{Sec:Conclusions}, we show some computational results for $p=3$ and $p=5$, which point out that, unlike $\Z_2\Z_4$-linear GH codes, when $p\geq 3$ prime, the $\Z_{p^2}$-linear GH codes are not included in the family of $\Z_p\Z_{p^2}$-linear GH codes with $\alpha_1\not =0$. Moreover, we also observe that they are not equivalent to any of the  $\Z_{p^s}$-linear GH codes considered in \cite{KernelZ2s,HadamardZps} by using the same Gray map.
\section{Preliminary results}\label{Sec:GrayMap}

In this section, we give the definition of the Gray map considered in this paper for elements of $\Z_{p^2}$. We also include some of its properties used in the paper.

We consider the following Gray map $\phi$, given in \cite{Carlet,GrayIsometry}, for $s=2$:  
\begin{equation}
\begin{split}\label{eq:GrayMapCarlet2}
\phi: & \ \Z_{p^2} \longrightarrow \Z_p^p \\
      & \ u  \mapsto (u_0,u_1)M,
\end{split}
\end{equation}
where $u\in\Z_{p^2}$; $[u_0,u_1]_p$ is the $p$-ary expansion of $u$, that is, $u=u_0 + u_1p$ with $u_0, u_1 \in \Z_p$; and $M$ is the following matrix of size $2 \times p$:
$$\left(\begin{array}{ccccc}
0 & 1 & 2 &\cdots  & p-1 \\
1 & 1 & 1 &\cdots  & 1 \\
\end{array}\right).$$

Let $u,v\in\Z_{p^2}$ and $[u_0,u_1]_p$, $[v_0,v_1]_p$ be the
$p$-ary expansions of $u$ and $v$, respectively, i.e. $u=u_0+u_1p$ and $v=v_0+v_1p$. We define the operation ``$\odot_p$''  between elements in $\Z_{p^2}$ as $u\odot_p v=t_0+t_1p$, where 
$$
t_i=\left\lbrace\begin{array}{ll}
 1, & \textrm{if } \quad u_i+v_i \geq p,\\
0, &  \textrm{otherwise}.\\
\end{array}\right.
$$
Note that the $p$-ary expansion of  $u\odot_p v$ is $[t_0,t_1]_p$, where $t_0, t_1 \in \{0,1\}$.
We denote in the same way,  ``$\odot_p$'', the component-wise operation. For $\uu=(u\mid u'), \vv=(v\mid v') \in \Z_p^{\alpha_1} \times \Z_{p^2}^{\alpha_2}$, we denote $\uu \odot_p \vv=(\zero \mid u' \odot_p v')$.

From \cite{HadamardZps}, we have the following results: 

\begin{lemma}\cite{HadamardZps}\label{lemma1}
Let $u \in \Z_{p^2}$ and $\lambda \in \Z_p$. Then, 
$\phi(u+\lambda p)= \phi(u)+ (\lambda, \lambda ,\dots,\lambda)$.
\end{lemma}

\begin{corollary}\cite{HadamardZps}\label{coro4}
 Let $\lambda, \mu \in \Z_p$. Then,  $\phi(\lambda \mu p)= \lambda \phi(\mu p)=\lambda \mu \phi(p)$.
\end{corollary}

\begin{corollary} \cite{HadamardZps}\label{coro1}
Let $u, v$ $\in \Z_{p^2}$. Then, $\phi (u) + \phi (v) = \phi (u+v-p(u\odot_p v))$.
\end{corollary}

\begin{corollary} \cite{HadamardZps}\label{lemma2}
Let $u, v$ $\in \Z_{p^2}$. Then, $\phi(pu+v)= \phi(pu)+\phi(v)$.
\end{corollary}

\begin{corollary}\cite{HadamardZps}\label{coroNew}
  For $u, v$ $\in \Z_{p^2}$, $\phi (u+v)= \phi (u) + \phi (v) + (t_0,t_0,\dots,t_0)$, where $t_0= 1$ if $u_0+v_0 \geq p$ and $0$ otherwise.
\end{corollary}
\begin{proposition} \cite{HadamardZps}\label{GHpro}
Let $u,v\in\Z_{p^2}$ be two distinct elements. Then, $\phi(u) -\phi(v)=\phi(u-v)=(\lambda, \ldots, \lambda)$ if $u-v=\lambda p \in   p\Z_{p^2}$, and $\phi(u) -\phi(v)$ contains every element of $\Z_p$ exactly once if $u-v \in  \Z_{p^2} \setminus  p\Z_{p^2}$.
\end{proposition}

\begin{proposition} \cite{HadamardZps}\label{disweight}
Let $u,v\in\Z_{p^2}$. Then, $d_H(\phi(u),\phi(v))=\wt_H(\phi(u-v))$.
\end{proposition}

From \cite{Constantinescu}, the homogeneous weight of 
an element $u\in\Z_{p^2}$ is defined by
\begin{equation}\label{weight}
    \wt^*(u)=\left\lbrace\begin{array}{ll}
 0 & \textrm{if} \quad u=0, \\
p & \textrm{if} \quad u \in p\Z_{p^2}\setminus \{0\},\\
p-1 & \textrm{otherwise},\\
\end{array}\right.
\end{equation}
and the  homogeneous weight of a vector $\mathbf{u}=(u_1,\dots,u_n) \in \Z_{p^2}^n$ is $\wt^*(\mathbf{u})=\sum_{i=1}^{n}\wt^*(u_i)$.
The corresponding homogeneous distance of $\mathbf{u}=(u_1,\dots,u_n)$ and $\mathbf{v}=(v_1,\dots,v_n)  \in \Z_{p^2}^n$ is defined as follows:
\begin{equation}\label{LeeDistance}
   d^*(\mathbf{u}, \mathbf{v})= \sum_{i=1}^n \wt^*(u_i-v_i).
\end{equation}
The Gray map $\Phi$ over $\Z_{p^2}^n$ is an isometry which transforms homogeneous 
distances defined in $\Z_{p^2}^n$ to Hamming distances defined in $\Z_p^{np}$ \cite{GrayIsometry}.

Then, we define the homogeneous weight of $\mathbf{u}=(u\mid u')=(u_1,\dots,u_{\alpha_1}\mid u'_1,\dots,u'_{\alpha_2}) \in \Z_p^{\alpha_1} \times \Z_{p^2}^{\alpha_2} $ as
$\wt^*(\mathbf{u})=\wt_H(u)+ \wt^*(u')$.
From (\ref{LeeDistance}), the corresponding homogeneous distance of $\mathbf{u}=(u\mid u')=(u_1,\dots,u_{\alpha_1}\mid u'_1,\dots,u'_{\alpha_2})$ and $\mathbf{v}=(v\mid v')=(v_1,\dots,v_{\alpha_1}\mid v'_1, \dots,v'_{\alpha_2}) \in \Z_p^{\alpha_1} \times \Z_{p^2}^{\alpha_2}$ is defined as follows:
\begin{equation}\label{distance}
   d^*(\mathbf{u}, \mathbf{v})= \wt_H(u-v)+ \wt^*(u'-v').
\end{equation}
Note that the extension of the Gray map $\Phi$ over  $\Z_p^{\alpha_1} \times \Z_{p^2}^{\alpha_2}$ 
is also an isometry by using this homogeneous metric, that is, $d^*(\mathbf{u}, \mathbf{v})=d_H(\Phi(\mathbf{u}), \Phi(\mathbf{v}))$ for all $\mathbf{u}, \mathbf{v}\in \Z_p^{\alpha_1} \times \Z_{p^2}^{\alpha_2}$. Moreover, the $\Z_p\Z_{p^2}$-linear codes obtained from this Gray map $\Phi$ are distance invariant by Proposition \ref{disweight}.

\section{Construction of $\Z_p \Z_{p^2}$-additive GH codes}\label{Sec:construction}

The description of a generator matrix having minimum number of rows for
$\Z_2\Z_4$-additive GH codes with $\alpha_1\not =0$, as long as an iterative construction of these matrices, are given in \cite{PRV06,RSV08}. In this section, we generalize these results for $\Z_p\Z_{p^2}$-additive GH codes with $\alpha_1\not =0$ and any $p\geq 3$ prime. Specifically, we define an iterative construction for the generator matrices and establish that they generate $\Z_p\Z_{p^2}$-additive GH codes. The proof that the codes are GH is completely different from the binary case.

Let  $\mathbf{0}, \mathbf{1},\mathbf{2},\ldots, \mathbf{p^{2}-1}$ be the vectors having the elements $0, 1, 2, \ldots, p^2-1$  repeated in each coordinate, respectively. Let
\begin{equation}\label{eq:recGenMatrix0}
A_p^{1,1}=
\left(\begin{array}{cccc|ccccc}
1 & 1 &\cdots & 1  &p &p &\cdots &p \\
0  & 1 & \cdots & p-1  &1 &2 &\cdots &p-1  \\
\end{array}\right).
\end{equation}
Any matrix $A_p^{t_1,t_2}$ with $t_1\geq 1, t_2 \geq 2$ or $t_1\geq 2, t_2 \geq 1$ can be obtained by applying the
following iterative construction. First, if $A$ is a generator matrix of a $\Z_p\Z_{p^2}$-additive code, that is, a subgroup of $\Z_p^{\alpha_1} \times \Z_{p^2}^{\alpha_2}$, then we denote by $A_1$ the submatrix of $A$ with the first $\alpha_1$ columns over $\Z_p$, and $A_2$ the submatrix with the last $\alpha_2$ columns over $\Z_{p^2}$. We start with $A_p^{1,1}$. Then, if we have a matrix $A=A_p^{t_1,t_2}$,
we may construct the matrices 
\begin{equation}\label{eq:ZpZp2recGenMatrix2}
A_p^{t_1,t_2+1}=
\left(\begin{array}{cccc|cccc}
A_1 & A_1 &\cdots & A_1  &A_2 &A_2 &\cdots &A_2 \\
\mathbf{0}  & \mathbf{1} &\cdots & \mathbf{p-1} &p\cdot \mathbf{0} &p\cdot \mathbf{1} &\cdots &p\cdot \mathbf{(p-1)}  \\
\end{array}\right)
\end{equation}
and
\begin{equation}\label{eq:ZpZp2recGenMatrix1}
\footnotesize
A_p^{t_1+1,t_2}=
\left(\begin{array}{cccc|ccccccc}
A_1 & A_1 &\cdots & A_1  &pA_1 &\cdots &pA_1 & A_2 &A_2&\cdots &A_2 \\
\mathbf{0}  & \mathbf{1} &\cdots & \mathbf{p-1}  &\mathbf{1} &\cdots &\mathbf{p-1} & \mathbf{0} &\mathbf{1}&\cdots &\mathbf{p^2-1}  \\
\end{array}\right).
\end{equation}

\begin{example}\label{Ex:11}
 Let
$$
A_3^{1,1}=
\left(\begin{array}{ccc|cc}
1 & 1 & 1  &3 &3 \\
0  & 1 & 2  &1 &2  \\
\end{array}\right)
$$
be the matrix described in (\ref{eq:recGenMatrix0}) for $p=3$. By using the constructions described in (\ref{eq:ZpZp2recGenMatrix2}) and (\ref{eq:ZpZp2recGenMatrix1}), we obtain $A_3^{1,2}$ and $A_3^{2,1}$, respectively, as follows:
$$
A_3^{1,2}=
\left(\begin{array}{ccccccccc|cccccc}
1 &1 &1 &1 &1 &1 &1 &1 &1 &3 &3 &3 &3 &3 &3 \\
0 &1 &2 &0 &1 &2 &0 &1 &2 &1 &2 &1 &2 &1 &2 \\
0 &0 &0 &1 &1 &1 &2 &2 &2 &0 &0 &3 &3 &6 &6\\
\end{array}\right)
$$
$$
\footnotesize
A_3^{2,1}=
\left(\begin{array}{ccccccccc|ccccccccccccc}
 1 &1 &1 &1 &1 &1 &1 &1 &1 &3 &3 &3 &3 &3  &3 &3 &3 &3 &3 &\cdots &3 &3\\
 0 &1 &2 &0 &1 &2 &0 &1 &2 &0 &3 &6 &0 &3 &6  &1 &2 &1 &2 &\cdots &1 &2\\
 0 &0 &0 &1 &1 &1 &2 &2 &2 &1 &1 &1 &2 &2 &2 &0 &0 &1 &1 &\cdots &8 &8\\ 
\end{array}\right).
$$
\end{example}

In this paper, 
we consider that the matrices $A_p^{t_1,t_2}$ are constructed recursively starting from $A_p^{1,1}$ in the following way. First, we add $t_1-1$ rows of order $p^2$, up to obtain $A_p^{t_1,1}$; and then 
we add $t_2-1$ rows of order $p$ up to achieve $A_p^{t_1,t_2}$.  Note that in the first row there is always the row $(\one \mid \p)$.

The $\Z_p\Z_{p^2}$-additive code generated by $A_p^{t_1,t_2}$ is denoted by ${\cH}_p^{t_1,t_2}$, and the corresponding $\Z_p\Z_{p^2}$-linear code $\Phi( {\cH}_p^{t_1,t_2})$ by $H_p^{t_1,t_2}$. We also write $A^{t_1,t_2}$, ${\cH}^{t_1,t_2}$, and $H^{t_1,t_2}$ instead of $A_p^{t_1,t_2}$, ${\cH}_p^{t_1,t_2}$, and $H_p^{t_1,t_2}$, respectively, when the value of $p$ is clear by the context.

\begin{proposition} \label{lemm:length}
Let $t_1,t_2\geq1$ and $p$ prime. 
Then, ${\cH}_p^{t_1,t_2}$ is a $\Z_p\Z_{p^2}$-additive code  of type \begin{equation}\label{eq:type} (p^{t_1+t_2-1}, (p-1)\displaystyle\sum_{i=1}^{t_1}p^{t_1+t_2+i-3};t_1,t_2).
\end{equation}
\end{proposition}
\medskip

\begin{theorem}\label{Th:GH11}
The $\Z_p\Z_{p^2}$-additive code ${\cH}_p^{1,1}$ generated by the matrix
$$
A_p^{1,1}=
\left(\begin{array}{cccc|ccccc}
1 & 1 &\cdots & 1  &p &p &\cdots &p \\
0  & 1 & \cdots & p-1  &1 &2 &\cdots &p-1  \\
\end{array}\right)
$$
 is a $\Z_p\Z_{p^2}$-additive GH code of type $(p,p-1;1,1)$.
\end{theorem}

\begin{example}\label{Ex3_11}
The $\Z_3\Z_{9}$-additive code ${\cH}_3^{1,1}$  generated by the  matrix $A_3^{1,1}$, given in Example \ref{Ex:11},  
is a $\Z_3\Z_9$-additive GH code of type $(3,2;1,1)$. Indeed, we have that $H_3^{1,1}=\Phi({\cH}_3^{1,1} )= \bigcup_{\lambda \in\Z_3} (\Phi(A_0) + \lambda \textbf{1})$, where $A_0=\{\lambda(0,1,2\mid 1,2): \lambda \in \Z_9\}$, and then $\Phi(A_0)$ consists of all the rows of the GH matrix 
\begin{equation}
    H(3,3)=
    \left(\begin{array}{ccccccccc}
0 &0 &0 &0 &0 &0 &0 &0 &0 \\
0 &1 &2 &0 &1 &2 &0 &2 &1  \\
0 &2 &1 &0 &2 &1 &1 &2 &0 \\
0 &0 &0 &1 &1 &1 &2 &2 &2\\
0 &1 &2 &1 &2 &0 &2 &1 &0\\
0 &2 &1 &1 &0 &2 &0 &1 &2\\
0 &0 &0 &2 &2 &2 &1 &1 &1\\
0 &1 &2 &2 &0 &1 &1 &0 &2\\
0 &2 &1 &2 &1 &0 &2 &0 &1\\
\end{array}\right).
\end{equation}
The $\Z_3\Z_{9}$-linear code $H_3^{1,1}$ has length $N=9$, $pN=3 \cdot 9=27$ codewords and minimum distance $N(p-1)/p=9(3-1)/3=6$.

\end{example}

\begin{theorem}\label{Th:GH}
The $\Z_p\Z_{p^2}$-additive code $\mathcal{H}_p^{t_1,t_2}$ generated by the matrix $A_p^{t_1,t_2}$, with $t_1, t_2\geq 1$ and $p$ prime, is a $\Z_p\Z_{p^2}$-additive GH code.
\end{theorem}

\begin{proposition}
Let $\mathcal{H}_p^{t_1,t_2}$ be a $\Z_p\Z_{p^2}$-additive GH code of type $(\alpha_1, \alpha_2; t_1,t_2)$ with $t_1, t_2 \geq 1$ and $p$ prime. Let $H_p^{t_1,t_2}$ be the corresponding $\Z_p\Z_{p^2}$-linear  GH code of length $p^t$, with $t\geq 2$. Then, $\alpha_1=p^{t-t_1}$, $\alpha_ 2=p^{t-1}-p^{t-t_1-1}$ and $t=2t_1+t_2-1$.
\end{proposition}


\begin{example}
Let ${\cH}_3^{1,2}$ be the $\Z_3\Z_{9}$-additive code generated by the matrix $A_3^{1,2}$ given in Example \ref{Ex:11}. By Theorem \ref{Th:GH}, $H_3^{1,2}=\Phi({\cH}_3^{1,2} )$ is a $\Z_3\Z_{9}$-linear GH code of type $(9,6;1,2)$. Actually, we can write
 $H_3^{1,2}= \cup_{\lambda \in\Z_3} (F_H + \lambda \textbf{1})$, where $F_H$ consists of all the rows of a GH matrix $H(3,9)$. Also, note that  $H_3^{1,2}$ has length $N=27$, $pN=3 \cdot 27=81$ codewords and minimum distance $N(p-1)/p=27(3-1)/3=18$.
\end{example}

\begin{example}
Let ${\cH}_3^{2,1}$ be the $\Z_3\Z_{9}$-additive code generated by the matrix $A_3^{2,1}$ given in Example \ref{Ex:11}. By Theorem \ref{Th:GH}, $H_3^{2,1}=\Phi({\cH}_3^{2,1} )$ is a $\Z_3\Z_{9}$-linear GH code of type $(9,24;2,1)$, which has length $N=81$, $pN=3 \cdot 81=243$ codewords and minimum distance $N(p-1)/p=81(3-1)/3=54$.
\end{example}

\medskip
Let $\mathcal{H}$ be a $\Z_p\Z_{p^2}$-additive code of type $(\alpha_1, \alpha_2; t_1,t_2)$ with  $p$ prime.
Let $\mathcal{H}_1$ (respectively, $\mathcal{H}_2$) be the punctured code of $\mathcal{H}$ by deleting the last $\alpha_2$ coordinates over $\Z_{p^2}$  (respectively, the first $\alpha_1$ coordinates over $\Z_p$). Let $H_2=\Phi(\mathcal{H}_2 )$.

\begin{remark}\label{remark2} 
    Let $\mathcal{H}=\mathcal{H}_p^{t_1,t_2}$ be a $\Z_p\Z_{p^2}$-additive GH code of type $(\alpha_1, \alpha_2; t_1,t_2)$ with $t_1, t_2 \geq 1$ and $p$ prime. Let $H=\Phi(\mathcal{H}_p^{t_1,t_2})$ be the corresponding $\Z_p\Z_{p^2}$-linear GH code of length $\alpha_1+p\alpha_2$.  Then, since $H$ is a GH code, its minimum distance  is $$\frac{(p-1)(\alpha_1+p\alpha_2 )}{p}.$$ Note that, by construction, $\mathcal{H}_1$ is a GH code over $\Z_p$ of length $\alpha_1$ and  minimum distance $(p-1)\alpha_1/p$. Therefore, $H_{2}=\Phi(\mathcal{H}_{2})$ has minimum distance $(p-1)\alpha_2$. 
\end{remark}

\begin{remark} \label{Remark:MinDistH-1-1}
Since the length of the $\Z_p\Z_{p^2}$-linear GH code $\Phi(\mathcal{H}_p^{1,1})$ is $p^2$, its minimum distance is $(p-1)p^2/p=p(p-1)$ by Remark \ref{remark2}.
\end{remark}

\begin{remark} \label{Remark:Alpha2NonZero}
The above constructions (\ref{eq:ZpZp2recGenMatrix2}) and (\ref{eq:ZpZp2recGenMatrix1}) give always $\Z_p\Z_{p^2}$-linear GH codes with $\alpha_2\not =0$ since the starting matrix $A_p^{1,1}$ has $\alpha_2\not = 0$. If $\alpha_2=0$, the $\Z_p\Z_{p^2}$-linear GH codes coincide with the codes obtained from a Sylvester GH matrix, so they are always linear and of type $(p^{t_2-1},0;0,t_2)$ \cite{dougherty2015ranks}. Therefore, we only focus on the ones with $\alpha_2\not =0$ to study whether they are linear or not.  
\end{remark}

\section{Other equivalent $\Z_p\Z_{p^2}$-linear GH codes}

In this section, we see that if we consider other starting matrices, instead of the matrix $A_p^{1,1}$ given in (\ref{eq:recGenMatrix0}), and apply the same recursive constructions $(\ref{eq:ZpZp2recGenMatrix2})$ and $(\ref{eq:ZpZp2recGenMatrix1})$, we also obtain $\Z_p\Z_{p^2}$-additive GH codes with $\alpha_2\not =0$. Indeed, the corresponding $\Z_p\Z_{p^2}$-linear GH codes, after applying the Gray map $\Phi$, are permutation equivalent to the codes $\Phi(\mathcal{H}_p^{t_1,t_2})$ constructed in Section \ref{Sec:construction}.

Let $\cS_n$ be the symmetric group of permutations on the set $\{1,\dots,n\}$. A permutation $\pi \in \cS_n$ acts linearly on vectors $(c_1,\ldots,c_n)\in \Z_p^n$ by permuting their coordinates as follows: $\pi(c_1,\ldots,c_n)=(c_{\pi^{-1}(1)}, \ldots, c_{\pi^{-1}(n)} )$. Given two permutations $\pi_1 \in \cS_{n}$ and  $\pi_2 \in \cS_{m}$, we define $(\pi_1 | \pi_2) \in \cS_{n+m}$, where  $\pi_1$ acts on the coordinates $\{1, \dots, n\}$ and $\pi_2$ on $\{n+1, \dots, n+m\}$.

Two codes $C_1$ and $C_2$ over $\Z_p$ of length $n$ are said to be monomially equivalent (or just equivalent) provided there is a monomial matrix $M$ such that $C_2=\{ \textbf{c} M : \textbf{c} \in C_1 \}$. Recall that a monomial matrix is a square matrix with exactly one nonzero entry in each row and column. They are said to be permutation equivalent if there is a permutation matrix $P$ such that $C_2=\{ \textbf{c}P : \textbf{c} \in C_1\}$. Recall that a permutation matrix is a square matrix with exactly one 1 in each row and column and $0$s elsewhere. A permutation matrix represents a permutation of coordinates, so we can also say that they are permutation equivalent if there is a permutation of coordinates $\pi\in \cS_n$ such that $C_2=\{ \pi(\textbf{c}) : \textbf{c} \in C_1 \}$.

 We denote by $N_p$ the set $\{0,1,\dots,p-1\}\subset \Z_{p^2}$ and $N_p^-=N_p\setminus\{0\}$. When including all the elements in those sets as coordinates of a vector, we place them in increasing order. For example, $N_3=\{0,1,2\}\subset \Z_9$,  $N_3^-=\{1,2\}\subset \Z_9$.

\begin{proposition} \label{prop:case11}
Let $a=(a_1,\ldots,a_{p-1})\in \Z_{p^2}^{p-1}$ such that $\{pa_1, pa_2,$ $\dots,$ $pa_{p-1}\}=p\Z_{p^2} \setminus \{0\}$. 
Consider the matrix
\begin{equation}\label{eq:recGenMatrix00}
A_{p,a}^{1,1}=
\left(\begin{array}{cccc|ccccc}
1 & 1 &\cdots & 1  &p &p &\cdots &p \\
0  & 1 & \cdots & p-1  &a_1 &a_2 &\cdots &a_{p-1}  \\
\end{array}\right).
\end{equation}
The code generated by $A_{p,a}^{1,1}$, denoted by $\mathcal{H}_{p,a}^{1,1}$, is a $\Z_p\Z_{p^2}$-additive GH code of type $(p,p-1;1,1)$.

Moreover, the corresponding $\Z_p\Z_{p^2}$-linear code after the Gray map is permutation equivalent to $\Phi(\mathcal{H}_p^{1,1})=\Phi( \mathcal{H}_{p,N_p^-}^{1,1})$. 
\end{proposition}

\begin{theorem} 
Let $a=(a_1,\ldots,a_{p-1})\in \Z_{p^2}^{p-1}$ such that $\{pa_1, pa_2,$ $\dots,$ $pa_{p-1}\}=p\Z_{p^2} \setminus \{0\}$. Let $A_{p,a}^{t_1,t_2}$ be the matrix obtained by using constructions $(\ref{eq:ZpZp2recGenMatrix2})$ and $(\ref{eq:ZpZp2recGenMatrix1})$, starting with the following matrix
\begin{equation}\label{eq:recGenMatrix01}
A_{p,a}^{1,1}=
\left(\begin{array}{cccc|ccccc}
1 & 1 &\cdots & 1  &p &p &\cdots &p \\
0  & 1 & \cdots & p-1  &a_1 &a_2 &\cdots &a_{p-1}  \\
\end{array}\right),
\end{equation}
instead of $A_p^{1,1}$. Then, the codes generated by $A_{p,a}^{t_1,t_2}$, denoted by $\mathcal{H}_{p,a}^{t_1,t_2}$, are $\Z_p\Z_{p^2}$-additive GH codes of type $(\alpha_1,\alpha_2;t_1,t_2)$.

Moreover, the corresponding $\Z_p\Z_{p^2}$-linear codes after the Gray map are permutation equivalent to $\Phi(\mathcal{H}_p^{t_1,t_2})=\Phi( \mathcal{H}_{p,N_p^-}^{t_1,t_2})$. 
\end{theorem}

\begin{example} 
Let $p=3$. Let $S=\{(a_1,a_2): (3a_1,3a_2)=(3,6)= 3\Z_9\setminus \{0\}\}$. Note that $S=\{(1,2),(1,5),(1,8),(4,2),(4,5),(4,8),(7,2),(7,5),(7,8)\}$ and it can also be written as $\{(1+3x, 2+3y) : x,y \in N_3\}$. Therefore, in this case, there are $9$ different starting matrices 
\begin{equation*}
A_{3,a}^{1,1}=
\left(\begin{array}{ccc|cc}
1 & 1  & 1  &3 &3 \\
0  & 1 & 2  &a_1 &a_2  \\
\end{array}\right),
\end{equation*}
where $a=(a_1,a_2)\in S$. If $a=(1,2)$, we obtain the matrix $A_3^{1,1}$ given in Example \ref{Ex:11}. These $9$ matrices generate $9$ different $\Z_3\Z_{9}$-additive codes $\mathcal{H}_{3,a}^{1,1}$ whose corresponding $\Z_3\Z_9$-linear codes, $\Phi(\mathcal{H}_{3,a}^{1,1})$, are permutation equivalent to each other. Moreover, note that if we permute the coordinates of $a$, we also obtain  $\Z_3\Z_{9}$-linear codes which are permutation equivalent to the previous ones.
\end{example}

\section{Linearity of $\Z_p\Z_{p^2}$-linear GH codes}\label{Sec:Linearity}

 In \cite{PRV06}, it is shown that the $\Z_2\Z_4$-linear GH codes of type $(\alpha_1,\alpha_2;1,t_2)$ are the only ones  which are linear, when $\alpha_1\not =0$. The next result shows that there are no $\Z_p\Z_{p^2}$-linear GH codes of type $(\alpha_1, \alpha_2; t_1,t_2)$, with $\alpha_1\not =0$, $t_1,t_2\geq 1$ and  $p\geq 3$ prime, which are linear. Note that this result for $p\geq 3$ does not coincide with the known result for $p=2$ if $t_1=1$. As it is mentioned in Remark \ref{Remark:Alpha2NonZero}, we only need to focus on codes with $\alpha_2\not =0$.

\begin{theorem}\label{th:linear}
    	Let $\mathcal{H}_p^{t_1,t_2}$ be the $\Z_p\Z_{p^2}$-additive GH code of type $(\alpha_1, \alpha_2; t_1,t_2)$ with $\alpha_1\not =0$, $t_1,t_2\geq 1$ and  $p\geq 3$ prime. Then, $H_p^{t_1,t_2}=\Phi(\mathcal{H}_p^{t_1,t_2})$ is non-linear.
\end{theorem}

\begin{example}z
Let $\cH_3^{1,1}$ be the $\Z_3\Z_{9}$-additive GH code of type $(3,2;1,1)$ considered in Example \ref{Ex3_11}. For example, we have that $\Phi(0,1,2 \mid 1,2) +\Phi(0,2,1 \mid 2,4)=(0,1,2,0,1,2,0,2,1)+(0,2,1,0,2,1,1,2,0)=(0,0,0,0,0,0,1,1,1) \notin \Phi(\cH_3^{1,1})$, so  $H_3^{1,1}=\Phi(\cH_3^{1,1} )$ is a non-linear code.
\end{example}

\begin{example}
Considering all the integer solutions with $t_1, t_2\geq 1$ of the equation
$5=2t_1+t_2-1$, we have that
the $\Z_p\Z_{p^2}$-linear GH codes with $\alpha_1\not =0$ of length $p^5$, $p\geq 3$ prime, are the following: $H_p^{1,4}$ and $H_p^{2,2}$. By Theorem \ref{th:linear}, we have that $H_p^{1,4}$ and $H_p^{2,2}$ are non-linear. 
\end{example} 

\begin{example}
Considering all the integer solutions with $t_1, t_2\geq 1$ of the equation
$6=2t_1+t_2-1$, we have that
the $\Z_p\Z_{p^2}$-linear GH codes with $\alpha_1\not =0$ of length $p^6$, $p\geq 3$ prime, are the following: $H_p^{1,5}$, $H_p^{2,3}$ and $H_p^{3,1}$. By Theorem \ref{th:linear}, we have that  $H_p^{1,5}$, $H_p^{2,3}$ and $H_p^{3,1}$ are non-linear. 
\end{example} 

\section{Conclusions}
\label{Sec:Conclusions}

Two structural properties of codes over $\Z_p$ are the rank and
dimension of the kernel. The rank of a code $C$ over $\Z_p$ is simply the
dimension of the linear span, $\langle C \rangle$,  of $C$.
The kernel of a code $C$ over $\Z_p$ is defined as
$\mathrm{K}(C)=\{\textbf{x}\in \Z_p^n : \textbf{x}+C=C \}$ \cite{BGH83,pKernel}. If the all-zero vector belongs to $C$,
then $\mathrm{K}(C)$ is a linear subcode of $C$.
Note also that if $C$ is linear, then $K(C)=C=\langle C \rangle$.
We denote the rank of $C$ as $r$ and the dimension of the kernel as $k$.
The parameters $(r,k)$ can be used to distinguish between non-equivalent codes, since equivalent ones have the same value of rank and dimension of the kernel.

In \cite{Kro:2001:Z4_Had_Perf,PRV06}, the rank and dimension of the kernel are used to classify $\Z_2\Z_4$-linear Hadamard codes with $\alpha_1=0$ and $\alpha_1 \not =0$, respectively. Moreover, it is also known that the family of $\Z_2\Z_4$-linear Hadamard codes with $\alpha_1\not =0$ includes the family of $\Z_2\Z_4$-linear Hadamard codes with $\alpha_1=0$ \cite{KV2015}, since each $\Z_2\Z_4$-linear Hadamard code with $\alpha_1 =0$ is  equivalent to a  $\Z_2\Z_4$-linear Hadamard code with $\alpha_1 \not =0$. The rank and dimension of the kernel have also been used to classify $\Z_{p^s}$-linear GH codes of length $p^t$, with $p$ prime \cite{KernelZ2s,fernandez2019mathbb,EquivZ2s,HadamardZps}. 

Table \ref{table:TypesP3} shows the type $(\alpha_1,\alpha_2; t_1,t_2)$ and parameters $(r,k)$ for all $\Z_3\Z_9$-linear GH codes of length $3^t$, with $\alpha_1\not =0$ and $2\leq t\leq 8$, considered in this paper. It also includes the type $(0,\alpha_2;t_1,t_2)$ and parameters $(r,k)$ for all $\Z_9$-linear GH codes of the same length.  

By looking at Table \ref{table:TypesP3}, we have that all $\Z_3\Z_9$-linear GH codes of length $3^t$, with $\alpha_1\not =0$ and $2\leq t\leq 8$, are pairwise non-equivalent since all of them have a different value of the kernel. This means that there are at least $\lfloor t/2 \rfloor +1$ such codes for $2\leq t\leq 8$. Moreover, we can see that these non-linear codes are also non-equivalent to any $\Z_9$-linear GH code of the same length. Similar results can be obtained computationally for $p=5$ and $p=7$. This means that, unlike for $p=2$, in general, for $p\geq 3$ prime, the $\Z_{p^2}$-linear GH codes are not included in the family of $\Z_p\Z_{p^2}$-linear GH codes with $\alpha_1\not =0$. Finally, we can also observe that the $\Z_3\Z_9$-linear GH codes with $\alpha_1\not =0$ are not equivalent to any $\Z_{3^s}$-linear code of the same length, by comparing the values of $(r,k)$ given in Table  \ref{table:TypesP3} with the ones given in \cite[Tables 4 and 5]{HadamardZps}.

\begin{table}[ht!]
\centering
\begin{tabular}{|c|c||cc|cc|}
\cline{1-6}
 $p$ & $t$& \multicolumn{2}{c|}{$\Z_{9}$}& \multicolumn{2}{c|}{$\Z_3\Z_{9}$}\\
\cline{3-6}
&& $(0,\alpha_2;t_1,t_2)$ & $(r,k)$ & $(\alpha_1,\alpha_2; t_1,t_2)$ & $(r,k)$\\[0.2em]
\hline
$3$ & $2$  & (0,3;1,1)  &(3,3) &(9,0;0,3) &$(3,3)$\\&&&   &(3,2;1,1)    & $(4,2)$\\ 
\cline{2-6}
         &$3$                  &     $(0,9;1,2)$  & (4,4)            &$(27,0;0,4)$  &(4,4)  \\
         &&  $(0,9;2,0)$  & (5,2)            &$(9,6;1,2)$  &(5,3)  \\[0.2em]

\cline{2-6}
         &$4$                  &     $(0,27;1,3)$  & (5,5)  &$(81,0;0,5)$   &(5,5)\\        
         &&  $(0,27;2,1)$  & (6,3)         &$(27,18;1,3)$  &(6,4)\\
         &&& &$(9,24;2,1)$  &(10,3) \\[0.2em]
         
\cline{2-6}
         &$5$                  &     $(0,81;1,4)$  & (6,6)      &$(243,0;0,6)$    &(6,6)    \\
         &&  $(0,81;2,2)$  & (7,4)           &$(81,54;1,4)$  &(7,5)  \\
          &&  $(0,81;3,0)$  & (11,3)             &$(27,72;2,2)$  &(11,4)\\[0.2em]
          
\cline{2-6}
         &$6$                  &     $(0,243;1,5)$  & (7,7)         &$(729,0;0,7)$   &(7,7)  \\
         &&  $(0,243;2,3)$  & (8,5)          &$(243,162;1,5)$  &(8,6)  \\
          &&  $(0,243;3,1)$  & (12,4)            &$(81,216;2,3)$  &(12,5)\\
          &&& &$(27,234;3,1)$  &(20,4)\\[0.2em]  
          
 \cline{2-6}
         &$7$                  &     $(0,729;1,6)$  & (8,8)         &$(2187,0;0,8)$   &(8,8)  \\
         &&  $(0,729;2,4)$  & (9,6)          &$(729,486;1,6)$  &(9,7)   \\
          &&  $(0,729;3,2)$  & (13,5)          &$(243,648;2,4)$  &(13,6)   \\
           &&  $(0,729;4,0)$  & (21,4)            &$(81,702;3,2)$  &(21,5)\\[0.2em]  
           
 \cline{2-6}
         &$8$                  &     $(0,2187;1,7)$  & (9,9)         &$(6561,0;0,9)$   &(9,9)  \\
         &&  $(0,2187;2,5)$  & (10,7)          &$(2187,1458;1,7)$  &(10,8)   \\
          &&  $(0,2187;3,3)$  & (14,6)           &$(729,1944;2,5)$  &(14,7)  \\
           &&  $(0,2187;4,1)$  & (22,5)         &$(243,2106;3,3)$  &(22,6) \\
           &&& &$(81,2160;4,1)$  &(35,5)\\[0.2em]             
\hline
\end{tabular}
\caption{Type and parameters $(r,k)$ of $\Z_{9}$-linear  and $\Z_3\Z_9$-linear GH codes.}
\label{table:TypesP3}
\end{table}

\end{document}